\documentclass[]{aa}
\usepackage{multirow}
\usepackage{subfigure}
\usepackage{graphicx}
\usepackage{natbib}
\usepackage{appendix}
\usepackage{txfonts}
\titlerunning{Re-visiting the case of R Mon}
\authorrunning{Alonso-Albi, T. et al. 2018}

\begin{document} 

\newcommand\Msun{M$_{{\odot}}$\:}
\newcommand\pow[1]{10$^{#1}$}
\newcommand\Rin{R$_{\mathrm{in}}$\:}
\newcommand\Rout{R$_{\mathrm{out}}$\:}
\newcommand\kms{km~s$^{-1}$}
\newcommand\Lsun{L$_{{\odot}}$}
\newcommand\LEt{\textbf}

\title{Revisiting the case of R Monocerotis: Is CO removed at R$~<~$20~au?\footnote{The reduced PdBI datacubes are only available in electronic form at the CDS via anonymous ftp to cdsarc.u-strasbg.fr (130.79.128.5) or via http://cdsweb.u-strasbg.fr/cgi-bin/qcat?J/A+A/}}

   \subtitle{}

   \author{ T. Alonso-Albi\inst{1},
            P. Riviere-Marichalar\inst{2},
            A. Fuente\inst{1},
            S. Pacheco-V\'azquez\inst{1},
        B. Montesinos\inst{3},
            R. Bachiller\inst{1},
            \and
            S. P. Trevi\~no-Morales\inst{2,4}
            }

   \institute{
     Observatorio Astron\'omico Nacional (IGN), Calle Alfonso XII, 3, 28014 Madrid, Spain\\
          \email{t.alonso@oan.es,a.fuente@oan.es}
   \and
      Instituto de Ciencia de Materiales de Madrid (ICMM-CSIC). E-28049,
 Cantoblanco, Madrid, Spain.\\
      \email{pablo.riviere@csic.es}
   \and
    Departamento de Astrof\'isica, Centro de Astrobiolog\'ia (CAB, CSIC-INTA), ESAC Campus, Camino Bajo del Castillo s/n, 
    E-28692 Villanueva de la Ca\~{n}ada, Madrid, Spain.
    \and
    Chalmers University of Technology, Onsala Space Observatory, 439 92 Onsala, Sweden.
             }

   \date{Received 27 July 2017 / Accepted 20 May 2018}

\abstract
   { 
   To our knowledge, R Mon is the only B0 star in which a gaseous Keplerian disk has been detected.
   However, there is some controversy about the spectral type of R Mon. Some
   authors propose that it could be a later B8e star, where disks are more common.   
   }
   {
   Our goal is to re-evaluate the R Mon spectral type and characterize
   its protoplanetary disk. 
}
   {
   The spectral type of R Mon has been re-evaluated using the available continuum data and UVES emission lines. 
    We used a power-law disk model to fit previous $^{12}$CO 1$\rightarrow$0 and 2$\rightarrow$1 interferometric 
    observations and the PACS CO data to investigate the disk structure. 
    Interferometric detections of $^{13}$CO J=1$\rightarrow$0, HCO$^+$ 1$\rightarrow$0, and CN 1$\rightarrow$0 
    lines using the IRAM Plateau de Bure Interferometer (PdBI) are presented. The HCN 1$\rightarrow$0 line was not detected.
    }
   {
   Our analysis confirms that R Mon is a B0 star. The disk model compatible with the 
   $^{12}$CO 1$\rightarrow$0 and 2$\rightarrow$1 interferometric observations falls short of predicting the observed fluxes of 
   the 14$<$J$_u$$<$31 PACS lines; this is consistent with the scenario in which some contribution to these lines is coming from a 
   warm envelope and/or UV-illuminated outflow walls. More interestingly, the upper limits to the fluxes of 
   the J$_u$$>$31 CO lines suggest the existence of a region empty of CO at R$\lesssim$20 au in the proto-planetary disk. 
   The intense emission of the HCO$^+$ and CN lines shows the strong influence of UV photons on gas chemistry.
   }
{ 
The observations gathered in this paper are consistent with the presence of a transition disk with a cavity of R$_{in}$$\gtrsim$20~au around R Mon. This size is similar to the photoevaporation radius that supports the interpretation that UV photoevaporation is  main disk dispersal mechanism in massive stars
}

    \maketitle

\section{Introduction}
Herbig Ae/Be stars are intermediate mass (M$\sim$2--8 M$_\odot$) pre-main
sequence objects. These objects share many characteristics
with high-mass stars such as clustering and the existence of photo-dissociation regions (PDRs) in their surroundings, but their study presents an important 
advantage: there are many located closer to the Sun (d $\leq$ 1 Kpc) and 
in regions less complex than massive star forming regions. 
The  detection and detailed study of circumstellar disks around Herbig Ae/Be stars 
could provide valuable information to understand the formation of massive stars. 

In recent years, there has been much theoretical and observational work with the aim of understanding disk occurrence and evolution in Herbig Ae/Be stars.
Similar to the case of their low-mass counterpart T Tauri stars (TTs), the spectral energy distribution (SED) of 
many of these stars is characterized by an infrared excess due to thermal emission of circumstellar
dust. But the exact geometry of this circumstellar dust (disk, disk+shell, shell) is not clear. It is generally accepted that disks similar to those associated with TTs surround Herbig Ae and late-type Be stars \citep{Alo09,Boissier2011}, but their existence around more massive Be stars (spectral type earlier than B6) is uncertain.
Evidence of the existence of dusty and gaseous disks around some Herbig Be stars exist at
optical, near-infrared (NIR), and mid-IR wavelengths \citep{Meeus2001, Vink2002, MillanGabet2001, Acke2005}, 
but the direct detection of the disks 
at millimeter wavelengths remains elusive with only a handful of detections \citep{Alo09}. 
Since the disk is optically thick at NIR and mid-IR wavelengths, millimeter
observations are required to estimate the dust and gas mass in the disk. On the 
other hand, spectroscopic observations of molecular lines are useful 
to study the disk dynamics.

Observations of disks have been carried out using the Photoconductor Array Camera and Spectrometer (PACS) instrument
onboard the Herschel Space Observatory (HSO), within the Key Programs Dust, Ice, and Gas In Time (DIGIT, \cite{green2013})
and Gas in Protoplanetary Systems (GASPS, \cite{dent2013}). These observations report the detection
of pure rotational high-J (J$_u$ $>$ 14) CO emission lines in these sources (e.g.,
\citealp{Sturm2010, vanKempen2010, Meeus2012, Meeus2013}). Thermochemical models of UV irradiated 
disks show that these lines are tracing the warm gas (T$_k$$>$300 K) located in intermediate
layers between the disk surface and the midplane at
intermediate distances from the star (10 to 50 au) 
(e.g., \citealp{Jonkheid2007, Gorti2008, Woitke2009, Kamp2010, 
Bruderer2012, Fedele2013}). However, the PACS spectra are spectrally and spatially unresolved
and their comparison with models involves some ambiguity. The detected high-J CO 
emission could arise in the hot inner envelopes and/or molecular outflows associated with these young stars instead of in the UV-irradiated disk.
\citet{Fedele2013} derived, for the first time, the radial gas temperature gradient 
in the disk associated with the Herbig Ae star HD 100546 based on the combination of
data from PACS and spectrally resolved CO spectra from the Heterodyne Instrument for the Far-Infrared (HIFI) onboard the HSO. In \cite{Fedele2016}, this kind of study was
extended to HD 97048, AS 205, Oph-IRS 2-48, S CrA, TW Hya, HD 100546, and HD 163296.

R Mon is a very young Herbig Be star located at a distance of 800 pc \citep{Cohen1985}.
It is associated with the reflection nebula NGC 2261 and has a 
T Tauri companion located 0.7$\arcsec$ NW \citep{Close1997}.
The star drives a prominent molecular outflow that excites the HH 39 object, located in a small dark cloud 7' N 
from R Mon \citep{Canto1981,Jones1982,Brugel1984}. 
Based on continuum interferometric observations, \citet{Fuente2003} detected a $\sim$0.01 M$_\odot$ circumstellar disk 
toward R Mon. Subsequent interferometric observations of the CO 1$\rightarrow$0 and 2$\rightarrow$1 molecular lines showed 
the existence of a large gaseous disk in Keplerian rotation around the star \citep{Fuente2006}. \citet{Alo09} carried out a 
complete modeling of the SED including interferometric measurements at millimeter and centimeter wavelengths. They 
derived a more accurate value of the dust mass (1.4 $\times$10$^{-4}$ M$_\odot$) and showed that grain growth has proceeded 
to sizes of $\sim$1 cm in the disk midplane. In addition to the compact disk (R$_{in}$=18~au, R$_{out}$=150$\pm$50~AU), a 
toroidal envelope of $\sim$0.8 M$_\odot$ (R$_{in}$=700~au, R$_{out}$=12000~au) was needed to fit the single-dish far-infrared (FIR) 
fluxes \citep{Alo09}. 

Some controversy exists about the spectral type of R Mon, which is crucial for the correct interpretation of the
existing data. Early studies by \citet{the94} classified R Mon as a B0 star. \cite{Close1997} also classified this object as 
B0 after a detailed discussion of the available photometry, extinction, and estimated bolometric 
luminosity. Later, \citet{Mora01} 
classified R Mon as a B8 III, which is consistent with a luminosity of $\sim$450 L$_\odot$ derived by \citet{Natta93}. 
\citet{Fuente2006} determined that the mass of the star is 8$\pm$1 M$_\odot$ based on the Keplerian disk rotation, 
which is consistent with a young B0 star. 
However, \citet{Sandell2011} suggested that the rotation curve could be contaminated by the outflow emission 
and R Mon could rather be a late-type Be star in which circumstellar disks are common. 
In order to interpret the millimeter and infrared data, a critical revision of the spectral type is necessary.
If confirmed, R Mon would be one of the closest B0 stars and the only one with a disk
detected in molecular lines.  

In this paper, we present  new interferometric detections of the $^{13}$CO 1$\rightarrow$0, 
CN 1$\rightarrow$0, and HCO$^+$ 1$\rightarrow$0 lines toward R Mon. We revise the stellar spectral
type to determine the disk parameters taking advantage of the current available 
(PACS, interferometric) data. In more detail, in Sect. \ref{section:spectralType} we redo 
the determination of the R Mon spectral type using the stellar SED and high resolution spectroscopy in ultraviolet (UV) wavelengths. In Sect. \ref{section:Modelization} we investigate its physical structure 
based on the Herschel/PACS observations of the high-J CO lines combined with our interferometric CO 1$\rightarrow$0 and
2$\rightarrow$1 observations. Finally, we compare the new $^{13}$CO 1$\rightarrow$0, 
CN 1$\rightarrow$0 and HCO$^+$ 1$\rightarrow$0 lines with our model in Sect. \ref{section:Chem}.

\begin{figure}
    \centering
    \includegraphics[width=0.49\textwidth]{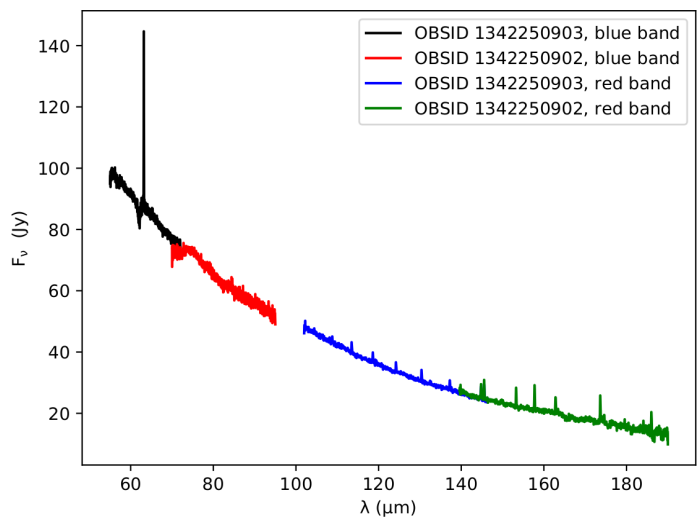}
    \caption {PACS spectroscopic observations from 50 to 190 $\mu$m, showing 
    the prominent [OI] emission at 63 $\mu$m as well as the CO lines.}
    \label{fig:PACSObs}%
\end{figure}

\begin{table}[]
\caption{CO line fluxes and widths derived from PACS observations at the central spatial pixel (spaxel).}           
\label{table:PACS_CO_fluxes} 
\centering                      
\begin{tabular}{crrr}       
\hline\hline
Transition       & \multicolumn{1}{c}{$\lambda$}         & \multicolumn{1}{c}{F} & \multicolumn{1}{c}{$\Delta$v}\\
                      & \multicolumn{1}{c}{($\rm \mu m$)}    & \multicolumn{1}{c}{$\rm (10^{-17} W m^{-2}$)} &  
                      \multicolumn{1}{c}{$\rm (km s^{-1})$)}\\ 
\hline
[OI]   $\rm ^{3}P_{1} \rightarrow ^{3}P_{2}$ & 63.185 & 250 $\rm \pm$ 9\\
\hline
CO 47$\rightarrow$46 & 56.121 & $\rm <$5.8  & 111 \\
CO 46$\rightarrow$45 & 57.307 & $\rm <$4.8  & 107  \\
CO 45$\rightarrow$44 & 58.547 & $\rm <$3.0  & 103 \\
CO 44$\rightarrow$43 & 59.843 & $\rm <$3.0  &  98  \\
CO 43$\rightarrow$42 & 61.201 & $\rm <$3.6  &  94 \\
CO 42$\rightarrow$41 & 62.624 & $\rm <$5.5  &  93 \\
CO 41$\rightarrow$40 & 64.117 & $\rm <$2.9  &  84  \\
CO 40$\rightarrow$39 & 65.686 & $\rm <$2.0  &  79  \\
CO 39$\rightarrow$38 & 67.366 & $\rm <$1.19 &  73 \\
CO 38$\rightarrow$37 & 69.074 & $\rm <$1.5 &  67 \\
CO 37$\rightarrow$36 & 70.907 & $\rm <$1.5 &  59  \\ 
CO 35$\rightarrow$34 & 74.890 & $\rm <$3.7  & 156   \\ 
CO 34$\rightarrow$33 & 77.059 & 3.9 $\rm \pm$ 0.7 & 151  \\
CO 33$\rightarrow$32 & 79.360 &  $\rm <$4.9 &  146 \\ 
CO 32$\rightarrow$31 & 81.806 & $\rm <$3.7  &  140 \\ 
CO 31$\rightarrow$30 & 84.410 & 6.7 $\rm \pm$ 1.1 $^{(1)}$  & 134 \\
CO 30$\rightarrow$29 & 87.190 &  3.2 $\rm \pm$ 0.7  & 127 \\
CO 29$\rightarrow$28 & 90.163 &  3.5 $\rm \pm$ 0.6  & 120 \\
CO 28$\rightarrow$27 & 93.349 &  6.3 $\rm \pm$ 1.8  & 113  \\
CO 25$\rightarrow$24 & 104.445 & 2.4 $\rm \pm$ 0.5 $^{(2)}$ & 319   \\
CO 24$\rightarrow$23 & 108.763 & 5.8 $\rm \pm$ 1.2  & 312  \\
CO 23$\rightarrow$22 & 113.458 & 11.0 $\rm \pm$ 1.0 $^{(1)}$ & 304    \\
CO 22$\rightarrow$21 & 118.581 & 7.4 $\rm \pm$ 0.6 & 296   \\
CO 21$\rightarrow$20 & 124.193 & 7.2 $\rm \pm$ 0.5  & 288 \\
CO 20$\rightarrow$19 & 130.369 & 7.6 $\rm \pm$ 0.6  & 279  \\
CO 19$\rightarrow$18 & 137.196 & 7.1 $\rm \pm$ 0.7  & 269 \\
CO 18$\rightarrow$17 & 144.784 & 7.6 $\rm \pm$ 0.5  & 258  \\
CO 17$\rightarrow$16 & 153.266 & 9.4 $\rm \pm$ 0.4 & 246   \\
CO 16$\rightarrow$15 & 162.812 & 7.7 $\rm \pm$ 0.6  & 233 \\
CO 15$\rightarrow$14 & 173.631 & 9.9 $\rm \pm$ 0.7  & 214 \\
CO 14$\rightarrow$13 & 185.199 & 8.5 $\rm \pm$ 1.3 & 194 \\
\hline                                   
\end{tabular}
\tablefoot{
$^{(1)}$ This line is blended with another transition and should be considered as an upper limit.
$^{(2)}$ This line is close to the edge of the spectrum, resulting in a feature that is too narrow. We characterize this as a lower limit.}
\end{table}

\begin{table*}[]
        \caption{Observational parameters related to the PdBI observations.}           
        \label{table:PdBI_parameters} 
        \centering                      
        \begin{tabular}{rrcrcrcc}       
        \hline\hline
Transition       &Frequency     &Beam              &PA        & Conversion factor  &  Noise     & Resolution  & Comments \\
              &(MHz)     &($\arcsec$)          &($^\circ$) &  (K beam mJy$^{-1}$)      &  (mJy beam$^{-1}$)  & (km~s$^{-1}$) \\ \hline
%
$^{13}$CO 1$\rightarrow$0 & 110201.354  &1.10$\times$0.66   &22       &  0.138    &  7.0     & 1.00 & Detection \\
$^{13}$CO 2$\rightarrow$1 & 220398.675  &0.61$\times$0.29  &20       &  0.142    & 12.0     & 1.00 & No detection \\
CN 1$\rightarrow$0       & 113490.985  &3.08$\times$2.90  &148      &  0.010     &  6.5     & 0.25 & Detection \\
HCO$^+$ 1$\rightarrow$0  &  89188.523  &5.19$\times$4.91 &119      &  0.006     &  4.0     & 0.30 & Detection \\
HCN 1$\rightarrow$0      &  88631.846  &5.23$\times$4.84 &125      &  0.006     &  4.8     & 0.30 & No detection \\
\hline                                   
\end{tabular}
\end{table*}

\section{Data acquisition}

\subsection{UVES/VLT spectrum}

We have used an archival spectrum from the UV and Visual Echelle Spectrograph (UVES) at Very Large Telescope (VLT), obtained on March
16, 2009. The data correspond to the reduced products provided by the ESO
archive. Although the two arms of UVES spectrograph overlap, the usable data
available have a gap between $\sim$4980--5680 \AA.

\subsection{Herschel - PACS}
R Mon was observed by the HSO in PACS range spectroscopic mode during September 2016, as 
part of the proposal \textit{$OT1\_gmeeus\_1$}. The $\rm \sim$ 50 $\rm \mu m$ to 190 $\rm \mu m$ wavelength range 
was covered by two different observations, with OBSID 1342250902 and 1342250903 (see Fig. \ref{fig:PACSObs}). The observations were performed 
in the pointed chop-nod mode to be able to subtract the telescope and background contributions. A small chopper 
throw (1.5$\arcsec$) was used, aiming to minimize the effect of field rotation between the two chop positions. 
Only one nodding cycle was performed for a total exposure time of 2240 s. The observations were reduced using 
HIPE 15, with the standard routine for chopped observations and the version 78 of the tree calibration files. In Table~\ref{table:PACS_CO_fluxes} and Table~\ref{table:PACSCO_H2O_fluxes}  we show the integrated line fluxes of 
the lines detected with PACS. Line fluxes were extracted from the central spatial pixel (spaxel) of the spectrometer, 
and aperture corrected to account for flux loss in the surrounding spaxels.

\subsection{Plateau de Bure Interferometer (PdBI)}

Interferometric observations of the $^{13}$CO 1$\rightarrow$0 and  2$\rightarrow$1 lines were carried out in January 2006 within
project P055. By that time the configuration of the receivers of the PdBI allowed us to observe the 3 mm and 1 mm bands simultaneously. 
The observations were performed in A configuration and six antennas. Correlator units of 80~MHz providing a spectral resolution of $\sim$78 kHz were used for the spectral observations of the $^{13}$CO 1$\rightarrow$0 and  2$\rightarrow$1 lines. The $^{13}$CO 1$\rightarrow$0 line was detected with a S/N ratio of $\sim$ 5 (integrated intensity). 
The $^{13}$CO 2$\rightarrow$1 line was not detected down to an rms of 12.0~mJy/beam in 1~km~s$^{-1}$ channel.  

The interferometric observations of CN, HCN, and HCO$^+$ were carried out in September (configuration D) and December 2008 (configuration C) 
(project S01A). The 80~MHz correlator unit was placed at 113.490~GHz providing a spectral resolution of 78~kHz to cover
the most intense hyperfine components of the CN 1$\rightarrow$0 line. 
The HCO$^+$ 1$\rightarrow$0 and HCN 1$\rightarrow$0 lines were observed simultaneously with the receiver centered at 
the frequency of 88.910~GHz. During the observations, two 40~MHz bandwidth correlator units 
were placed at the frequency of the HCN 1$\rightarrow$0 and HCO$^+$ 1$\rightarrow$0 lines, respectively,
providing a spectral resolution of $\sim$39~kHz (0.13~\kms). 
To improve the S/N ratio we degraded the velocity resolution of the HCO$^+$ and HCN lines 
to 0.3~\kms. The HCN 1$\rightarrow$0 line was not detected with a final rms of  $\sim$5~mJy/beam in a channel of 0.3~\kms.

Data reduction and image synthesis were carried out using the \texttt{GILDAS}\footnote{See \texttt{http://www.iram.fr/IRAMFR/GILDAS} 
for  more information about the GILDAS software~\citep{Pety2005}.} software. 
The channels free of line emission  were used to estimate the 
continuum flux that was subtracted from the spectral maps. 
The main observational parameters are listed in Table~\ref{table:PdBI_parameters}.

\begin{figure*}
    \centering
    \includegraphics[width=1.0\textwidth]{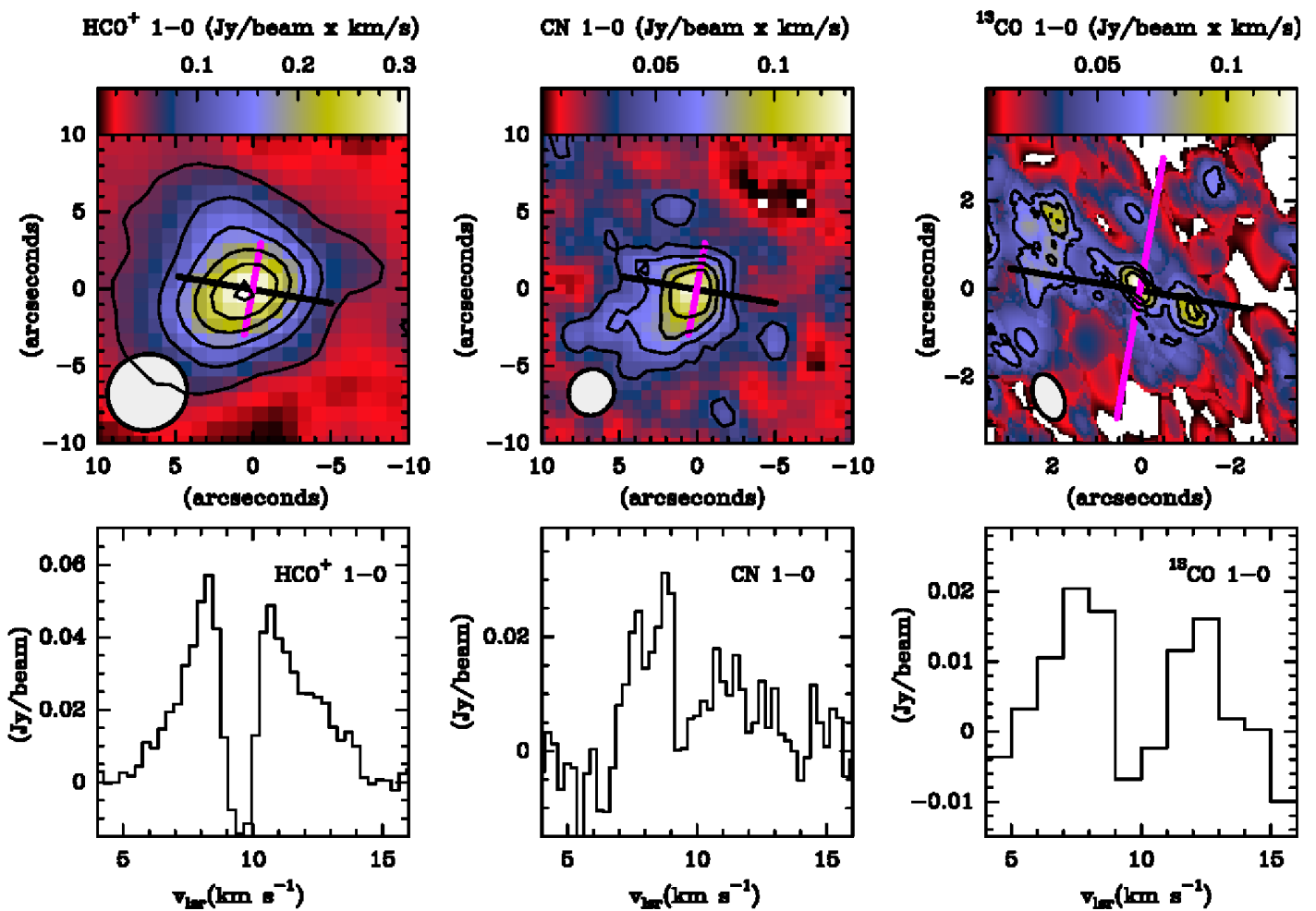}
    \caption{Integrated intensity maps of the HCO$^+$ J=1$\rightarrow$0, CN N=1$\rightarrow$0 (J,F=3/5,5/2$\rightarrow$1/3,3/2),
and $^{13}$CO J=1$\rightarrow$0 lines. The velocity range of the integration is from 5 to 15~km~s$^{-1}$. 
Contour levels are  0.05 to 0.3 in steps of 0.05 Jy/beam$\times$km~s$^{-1}$ for HCO$^+$; 0.04 to 0.12 in steps 
of 0.02 Jy/beam$\times$km~s$^{-1}$ for CN; and 0.05 to 0.3 in steps of 0.05 Jy/beam$\times$km~s$^{-1}$ for $^{13}$CO.
The orientation of the molecular emission is approximately perpendicular to the magenta arrow that indicates 
the direction of the outflow. The black line indicates the disk major axis orientation.
The ellipse in the left bottom corner indicates the synthesized beam size.}
    \label{fig:Maps}%
\end{figure*}

\begin{figure}
    \centering
    \includegraphics[width=0.5\textwidth]{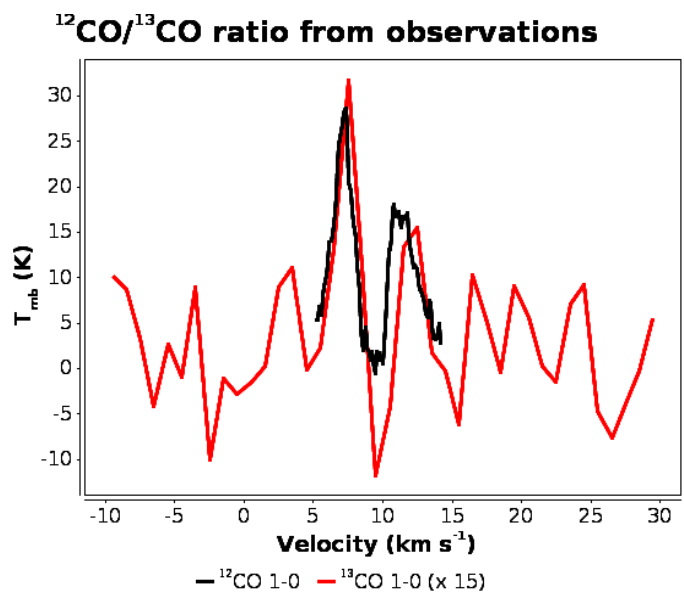}
   \caption{$^{12}$CO/$^{13}$CO ratio derived from IRAM Plateau de Bure observations.} 
    \label{fig:CO_13CO_ratio}%
\end{figure}

\begin{figure*}
    \centering
    \includegraphics[width=1.0\textwidth]{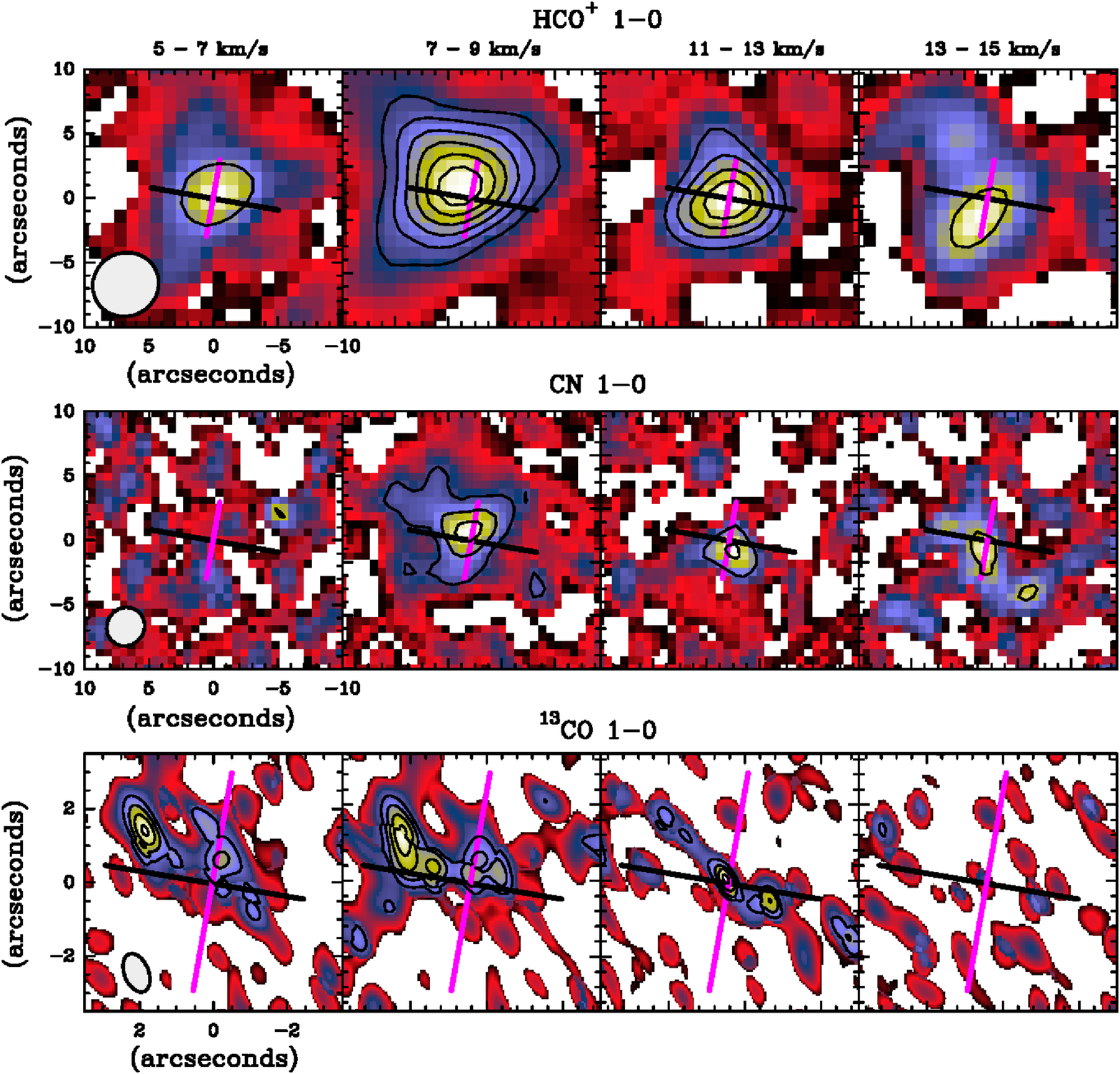}
   \caption{Integrated intensity maps of the HCO$^+$ J=1$\rightarrow$0, CN N=1$\rightarrow$0 (J,F=3/5,5/2$\rightarrow$1/3,3/2),
and $^{13}$CO J=1$\rightarrow$0 lines in various velocity intervals. We have avoided the central velocites (9$-$11~km~s$^{-1}$)
for which the space filtering effects are expected important. Contour levels are 0.03 to 0.4 in steps of 0.015~Jy$\times$km~s$^{-1}$ ({\it top});
0.02 to 0.12 in steps of 0.02~Jy$\times$km~s$^{-1}$ ({\it middle}); 0.03 to 0.12 in steps of 0.01~Jy$\times$km~s$^{-1}$ ({\it bottom}).
The orientation of the molecular emission is approximately perpendicular to the magenta arrow that indicates 
the direction of the outflow. The black line indicates the disk major axis orientation.
The ellipse in the left bottom corner indicates the synthesized beam size.} 
    \label{fig:Vel}%
\end{figure*}

\section{Observational results}

\subsection{PACS fluxes}

A variety of CO, $\rm H_{2}O,$ and OI emission lines were detected in the 50-200 $\rm \mu m$ range.  For
the detected lines, we computed line fluxes by fitting a Gaussian profile to the continuum subtracted spectra in a region one $\rm 
\mu m$ wide around each feature. The line flux uncertainties were estimated as a Gaussian integral with a peak equal 
to the noise of the continuum in the region of interest, and a width equal to the fitted value. For nondetected 
lines, we computed upper limits in a similar way as the integral of a Gaussian with a peak equal to the noise level 
of the continuum and a width equal to the instrumental value. 

The most prominent emission line detected was [OI] $^{3}P_{1} \rightarrow ^{3}P_{2}$ transition at 63.185 $\mu$m, 
which has a flux of  $\rm  (2.50 \pm 0.09) \times 10^{-15}$ W m$^{-2}$ (see Fig. \ref{fig:PACS_OImaps}), in
agreement with the value of $\rm  (2.37 \pm 0.08) \times 10^{-15}$ W m$^{-2}$ found by \cite{Riviere2016}. 
This line presents a complex profile with high velocity components at +137/-219 km~s$^{-1}$ very likely 
associated with the bipolar outflow.
The CO emission lines were detected for $\rm J_{u}$ in the range 31 to 14 and have fluxes in the range $\rm 3.2 \times 10^{-17}$ 
to $\rm 1.1 \times 10^{-16}$ $\rm W m^{-2}$ (see Table \ref{table:PACS_CO_fluxes}). Two CO transitions, J = 31$\rightarrow$30 at 
84.41 $\rm \mu m$ and J = 23$\rightarrow$22 at 113.458 $\rm \mu m$, are blended with OH and $\rm H_{2}O$ transitions, respectively, 
hence the fluxes that are computed have to be considered as upper limits. The CO J = 25$\rightarrow$24 transition at 104.445 
$\rm \mu m$ is located close to the edges of the spectrum, resulting in a too narrow line. Therefore, the flux reported has to be 
considered as a lower limit.

We applied to all CO lines the 
multi-Gaussian analysis described by \cite{Riviere2016}. In all the cases, the lines were better fitted with one single velocity component (see Fig. \ref{fig:PACS_detectedLines}).
This discards the existence of high velocity wings ($>$60 km~s$^{-1}$) in the J$_u$$>$30 CO lines (Table~\ref{table:PACS_CO_fluxes}).
Following the same methodology, line fluxes and upper limits for $\rm H_{2}O$ transitions are given in Table \ref{table:PACSCO_H2O_fluxes}.
The $\rm H_{2}O$ emission was detected at 108.073, 125.354, 179.525, and 180.487 $\rm \mu m$, in all cases with $\rm E_{u}/k_{B}<400$ K. 
But these detections are at a level of $<$5$\sigma$ (see Fig. \ref{fig:PACS_H2Omaps}), and taking into account the
uncertainties in the baselines, we consider the detection of $\rm H_{2}O$ is tentative.

Taking advantage of the spatial resolution provided by PACS we represented the line emission 
maps for the different species found in the wavelength range covered. Line fluxes were integrated in windows 
1 $\rm \mu m$ wide, and the continuum was computed in windows of similar size close to each line, in which no 
transition was present. The 25 spaxel were smoothed to 300$\rm \times$300 spatial resolution units. The maps for CO lines are 
shown in Fig. \ref{fig:PACS_COmaps_1} and \ref{fig:PACS_COmaps_2}. We performed tests to detect extended emission as in 
\cite{Podio2012}, but we did not detect extended emission for any of the observed [OI] and CO transitions up to the 3$\sigma$ 
level.

\subsection{Interferometric images}

In Fig. \ref{fig:Maps}, we show the integrated intensity maps of the HCO$^+$ J=1$\rightarrow$0, 
CN N=1$\rightarrow$0 J,F=3/5, 5/2$\rightarrow$1/3,3/2, and $^{13}$CO J=1$\rightarrow$0 lines obtained with the PdBI. 
The F=5/2$\rightarrow$3/2 line at 113490.97~MHz is the most intense hyperfine component of the CN N=1$\rightarrow$0 transition. 
We have not detected the hyperfine component at 113488.14~MHz with a 3$\times$rms upper limit of T$_b$(113488)/T$_b$(113491)$<$0.43. 
In the optically thin limit, the expected intensity of this line is 0.37 times that of the main component. Our limit is, 
thus, consistent with the CN emission being optically thin. No other hyperfine transition lies within the observed spectral range.
The profiles of the HCO$^+$ J=1$\rightarrow$0, 
CN N=1$\rightarrow$0 J,F=3/5, 5/2$\rightarrow$1/3,3/2, and $^{13}$CO J=1$\rightarrow$0 lines toward the star
position are shown in Fig.~\ref{fig:Maps}. 
We note that the low emission at the central velocities (9$-$11~km~s$^{-1}$) is due to spatial filtering
of the interferometer that misses the extended component.
For comparison, we show in Fig~\ref{fig:CO_13CO_ratio} the profile of $^{13}$CO 1$\rightarrow$0 line overlayed 
to the spectrum of the $^{12}$CO 1$\rightarrow$0 line reported by \citet{Fuente2006}.
Taking into account the limited S/N of the $^{13}$CO 1$\rightarrow$0 spectra, the velocity profiles of the two lines are very
consistent and T$_b$($^{12}$CO 1$\rightarrow$0)/T$_b$($^{13}$CO 1$\rightarrow$0)$\sim$15 (see Fig. \ref{fig:CO_13CO_ratio}). This value is 
$\sim$3 times larger than those derived in disks around less massive stars \citep{Dutrey1997, Pietu2007}, which suggests a 
warmer and less massive disk. 

The integrated intensity maps of the $^{13}$CO, HCO$^+$, and CN lines show an elongated structure 
with an orientation that is almost perpendicular to the outflow axis (PA$\sim$350$^\circ$) as determined by \cite{Brugel1984} from the [SII] jet
(see Fig.~\ref{fig:Maps}). Fig.~\ref{fig:Vel} shows the line integrated intensity 
for various velocity intervals avoiding the cloud systemic velocity. There is a clear velocity gradient
roughly perpendicular to the jet direction supporting the interpretation of a disk origin. Because of the larger beam of the HCO$^+$ and CN observations, we cannot discard some contribution from 
the envelope to the emission of these lines.

\section{UVES/VLT: Critical revision of the spectral type of R Mon}
\label{section:spectralType}

\begin{figure*}
    \centering
    \includegraphics[width=1.0\textwidth]{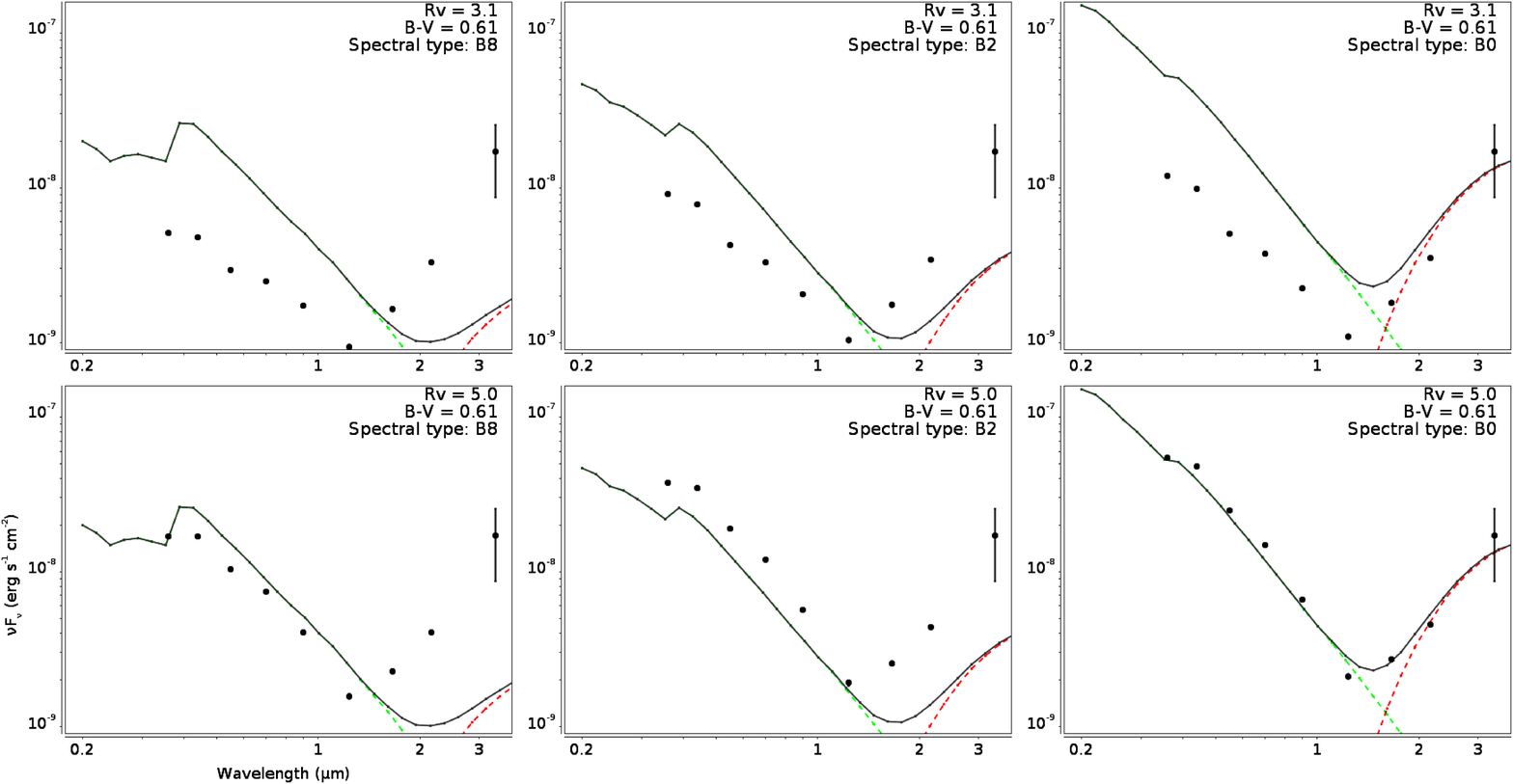}
    \caption{Fit to the UBVRI and JHK photometry assuming different values for the stellar parameters and extinction. The 
    green line indicates the stellar photospheric emission corrected for extinction. The red line is the dust thermal 
    emission coming from the disk. The gray line is the sum of both contributions.}
    \label{fig:RMonSpectralType}%
\end{figure*}

\begin{figure*}
    \centering
    \includegraphics[width=1.0\textwidth]{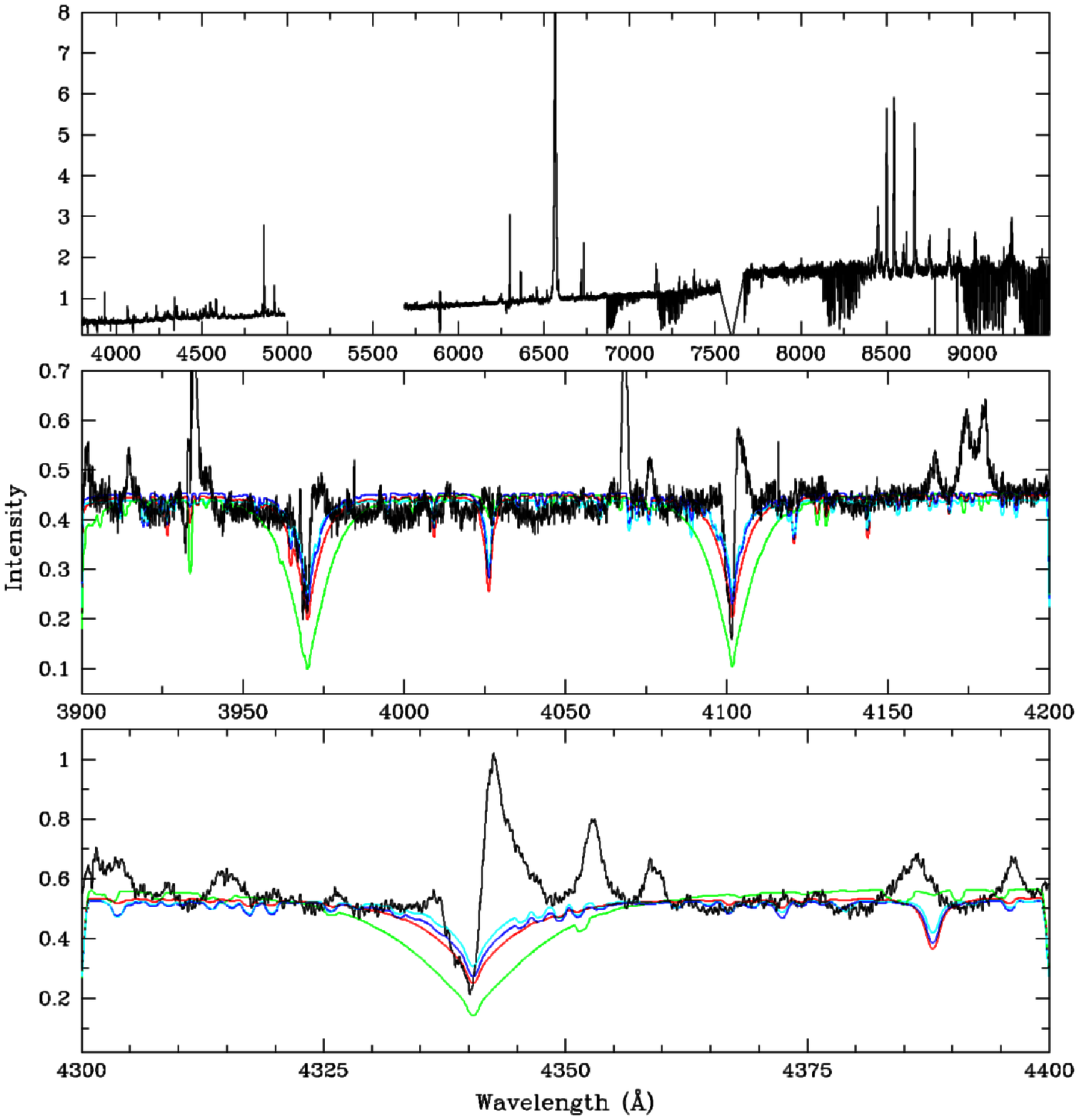}
    \caption{Spectrum of R Mon obtained with the UVES spectrograph on the
ESO/VLT. In the middle and bottom panels synthetic photospheric Kurucz
models with temperatures 11000, 20000, 25000, 30000 K are plotted in
green, red, blue, and cyan, respectively. Assuming that the absorption
features in the blue side of H$\epsilon$, H$\delta$, and H$\gamma$ have
a substantial photospheric contribution, the comparison with the
observations suggests that R Mon is an early B-type star with an effective 
temperature between 20000 and 30000 K.}
    \label{fig:RMonUVSpectra}%
\end{figure*}

We have estimated the spectral type of R Mon using both stellar  
SED and UVES emission lines.
The first problem when determining the correct spectral type of an embedded B star like R Mon, by fitting the continuum SED,
is the treatment of the extinction. As in \cite{Alo09}, we used the extinction law of \cite{Cardelli1989}.  
\citet{Her04} discussed the problem of the uncertainties in the extinction law and showed 
that in embedded Be stars the value of the total to selective absorption, R$_V$, is close to 
5, instead of the value of 3.1 commonly assumed for the dust in the galactic plane, due to the prevailing of a population of 
grains of greater size. The second problem is the possible variability of the source, which means that the observations should 
be obtained the same date. The latest photometry available for R Mon is the data reported by \cite{Mendi2012}, but the original source of the photometric data used by these authors is \citet{Morel1978}, which is 
the same set of fluxes adopted by \cite{Alo09}. Although old data, these UVBRI photometry were taken almost simultaneously so 
they are not affected by the source variability, showing a B-V color index of 0.61.  

Fig. \ref{fig:RMonSpectralType} shows our fits to the R Mon SED using various values of the spectral type.
We use the Kurucz model to predict the stellar photospheric emission. The near-IR excess 
can be explained by assuming the existence of a disk. At these wavelengths, the dust thermal emission is optically thick
and the flux mainly depends on the dust temperature in the disk surface that is determined by the spectral type
\citep{Goldreich97,Dullemond01}. We show the photospheric emission (green line) and 
the emission of the circumstellar disk that has been calculated according to the assumed spectral type, 
using the model described by \cite{Alo09} (red line). 
The UVBRI part of the SED as well as the near-IR excess can only be fitted by assuming R$_V$=5.0 and a spectral type B0 
for R Mon. A spectral type of B8 could only agree with the stellar photometry if we assume R$_V$ around 7, but this high 
value is very difficult to justify. 

An independent method to estimate the stellar spectral type is by fitting the UVES emission lines.
We plot the full UVES spectrum of R Mon in relative intensities in the top panel 
of Fig. \ref{fig:RMonUVSpectra}. The spectrum
is full of emission lines, of which the most prominent, from blue to red, are Ca {\sc K} 3933.66 \AA, H$\epsilon$, H$\delta$, H$\gamma$ (P
Cygni profiles), H$\beta$, S {\sc ii} 4925.35, Na {\sc i} 5889.95,
5895.92 \AA (both P Cygni), [O {\sc i}] 6300.23, 6363.88 \AA,
H$\alpha$, [N {\sc ii}] 6583.45 \AA\  (the nitrogen line at 6548.05 \AA\  
is blended with H$\alpha$), [S {\sc ii}] 6730.82, 6716.44 \AA, and
beyond 8500 \AA, the Ca {\sc ii} IR triplet (8495.02, 8542.09,
8662.14 \AA), and H {\sc i} Paschen 9--14 (only the completely certain
line identifications are listed). In addition, there are many
weak and broad features, also in emission, which are the result of
blends of many metallic lines. All these emissions are presumably
originated by the high accretion rate of this object (see below). The
absorptions in the regions beyond 6800, 7400, 8200, and 9000 \AA{} are
due to contamination by the Earth's atmosphere.

The middle and bottom panel of Fig. \ref{fig:RMonUVSpectra} show the
profiles of H$\epsilon$, H$\delta$, and H$\gamma$. As we pointed out above,
the three lines show what seem to be classical P Cygni profiles, which have
emission on the red side and absorption on the blue side. The blue wings in absorption of these three lines are virtually the
only features for which a comparison with photospheric synthetic
spectral models, in order to constrain the spectral type, is
feasible.

The programs {\sc synthe} and {\sc atlas} \citep{Kurucz1993} fed
with the models describing the stratification of the stellar
atmospheres \citep{Castelli2003} have been used for spectral
synthesis.  Models were computed for four temperatures, namely
11000 (spectral type $\sim$B8, plotted in green), 20000 (red), 25000
(blue), and 30000 K (cyan). 
It is clear that the profile with 11000 K deviates
most from the observations, whereas those in the interval 20000 --
30000 K (spectral type $\sim$B2--B0), although not fittting perfectly,
are much closer.
Solar abundances and values of $\log
g=4.0$ (typical of a hot star near the MS) and $v \sin i\!=\!50$
km/s were assumed. Given the total absence of photospheric
absorption metallic features it is impossible to estimate the
projected rotational velocity. The value assumed does not have a
significant effect on the results. As a test, the relative variation
of the widths of H$\gamma$ at intensity 0.8 with respect to a
continuum at 1.0, between two models with 20000 K and $v \sin
i\!=\!50$ and 200 km/s, is $\sim\!12$\%, therefore the results do not
change substantially if a larger rotation rate is used. The
synthetic photospheric spectra show absorption features that are
not seen in the spectrum of R Mon. This is the case of the prominent features He~{\sc
i} 4026.21, 4387.93 \AA~in the
narrow windows shown in Fig. \ref{fig:RMonUVSpectra}. A plausible explanation could be that the
high activity of the star manifested as emissions across the whole
spectrum erases those lines. Actually, two emissions slightly shifted
to the blue appear close to those two wavelengths.

 Concerning the accretion in R Mon, \cite{Fairlamb2015} (see
Sect. 6.1 of that paper) discussed this object and a few other
Herbig Be stars. R Mon presents a large blue excess around the Balmer
jump when compared with models where accretion is not included.
They are unable to give an accretion rate for R Mon within the
paradigm of the magnetospheric accretion model: it is not possible to
reproduce the large blue excess before the filling factor of accretion
shocks on the photosphere saturates, which implies that a different
scenario must be invoked for massive Be stars.

Considering the analysis in this section we conclude that R  Mon is more likely a B0 
star and this is the spectral type that we use hereafter.


\section{Disk model: $^{12}$CO and $^{13}$CO lines}
\label{section:Modelization}


Disk models with different complexity degrees are used to analyze
and interpret line and continuum observations from protoplanetary disks \citep{Jonkheid2007, Gorti2008, Woitke2009, Kamp2010, 
Bruderer2012, Fedele2013, Fedele2016}. These models are 
dependent on a wealth of parameters that are usually poorly known and for which one needs to assume a set of 
reasonable values. The disk mass, flaring, and density structure are key parameters 
to predict molecular emission. Other parameters such as dust settling, maximum grain size, grain size distribution 
and grain chemical composition, gas-to-dust mass ratio, and the PAH abundance, also have an impact on the chemistry, energetic
balance, and eventually on the predicted line emission (see, e.g., \citealp{Woitke2016, Fedele2016}). 

The CO lines are expected to be optically thick in most of the disk. 
Their intensities are hence determined by the gas temperature in the gas layer with $\tau \sim$1, which is close to 
the disk surface. Since the different CO rotational lines are excited at different gas kinetic temperatures, the SED
of the CO rotational lines is basically tracing the gas temperature profile on the disk surface.
For this reason, we decided to adopt a simple power-law model to fit the CO lines. 
In our approximation, the gas temperature and density
are described by the power-law profiles $T_{K}=T_{in} \times (R_{in} /R)^{q}$, $\Sigma\propto r^{-p}$. 
Even in this simple approximation the number of parameters is too large to be fitted only with the CO
observations. Thus, we fixed some of the parameters based on those previously derived from the continuum SED
\citep{Alo09}.

The gas mass is fixed to 0.014 M$_{\odot}$. This value is derived from the disk mass obtained by \citet{Alo09} 
and adopting a gas-to-dust ratio of 100. An outer radius, R$_{out}$=1500 au, was derived by \citet{Fuente2006}
by fitting the size of the compact emission detected in the $^{12}$CO 2$\rightarrow$1 line. Since this value is consistent with the size derived for the structure detected in the J, H, and K bands 
by \citet{Murakawa08}, we kept it in our new fitting. 
The $^{12}$CO abundance is
assumed uniform across the disk and equal to 1.6$\times$10$^{-4}$, which is a reasonable value for a disk in which the
gas has temperatures $>$25-30 K even in the midplane \citep{Alo10}. 
The adopted stellar 
parameters are T$_*$=25000 K and R$_*$=4 R$_\odot$, which are those derived
from the SED fitting shown in Fig. \ref{fig:RMonSpectralType} (R$_V$=5 and spectral type B0). 
Since we do not spatially resolve the disk even in the interferometric CO 1$\rightarrow$0 and 2$\rightarrow$1 
observations, the values of T$_{in}$ and R$_{in}$ cannot be derived independently. 
In our fitting, we assumed that the gas temperature in the inner edge
is equal to the blackbody limit, T$_{in}$= (R$_*$/R$_{in}$)$^{0.5}$ T$_*$, being R$_{in}$ 
a free parameter. 

Summarizing, in our procedure, we kept as fixed parameters M$_{d}$=0.014 M$_\sun$, X(CO)=1.6$\times$10$^{-4}$, and R$_{out}$=1500 au, and
varied R$_{in}$ (from 1 to 100 au, using the corresponding gas temperature T$_{in}$), $i$ (from 10 to 50 degrees), $p$ (from 0.5 to 0.7), 
$q$ (from 1.0 to 2.0) and the turbulent velociy $\sigma$ (from 0.1 to 5 km s$^{-1}$) to fit the observations. 
We used the ray-tracing radiative transfer code DATACUBE\footnote{This and other modeling 
tools used by our team can be installed following the instructions provided at conga.oan.es/\%7Ealonso/doku.php?id=jparsec.} 
to model the molecular emission coming from the disk.  
Using these parameters, we attempted to fit simultaneously our previous interferometric observations of the $^{12}$CO 1$\rightarrow$0 
and 2$\rightarrow$1 lines published by \citet{Fuente2006} and the PACS data. 
The interferometric profiles of the $^{12}$CO 1$\rightarrow$0 and 2$\rightarrow$1 lines are best fitted assuming  
Keplerian rotation around a 8$\pm$1 M$_\sun$ star and a turbulent dispersion of 0.5$\pm$0.3 km s$^{-1}$.
The power-law indexes are fited to $p$=0.60$\pm$0.1 and q=1.8$\pm$0.4 
(see Table \ref{table:model}). 
We note that these are the only spectrally resolved line profiles and then the
only information about the kinematical disk structure. The value of R$_{in}$ has a neglible influence
on the $^{12}$CO 1$\rightarrow$0 and 2$\rightarrow$1 line profiles because of the limited angular resolution 
and sensitivity of our observations. 

\begin{center}
\begin{table}
\caption{Fitting results for the CO emission}
\begin{tabular}{lc}\\  \hline \hline
\multicolumn{1}{c}{} &  \multicolumn{1}{c}{Flat disk} \\ \hline
R Mon mass (M$_\odot$)    &  8$\pm$1   \\
i ($^\circ$)                 &  20$\pm$10  \\
$R_{out}$  (au)                 &   1500     \\
$R_{in}$  (au)                 &   > 10     \\
$T_{in}$ (K) (r = 1 au)         &   3400$\pm$500  \\ 
$q$ (T power-law index)                        &   0.60$\pm$0.10 \\ 
$M_{disk}$ (M$_\odot$)           &  0.014$\pm$0.006 \\ 
$p$ ($\Sigma$ power-law index)              &    1.8$\pm$0.4   \\
$\Delta v_{turb}$  (km/s)  &   0.5$\pm$0.3  \\ 
\hline 
\label{table:model} 
\end{tabular}\\
\end{table}
\end{center}

\begin{figure*}
    \centering
    \includegraphics[width=1.0\textwidth]{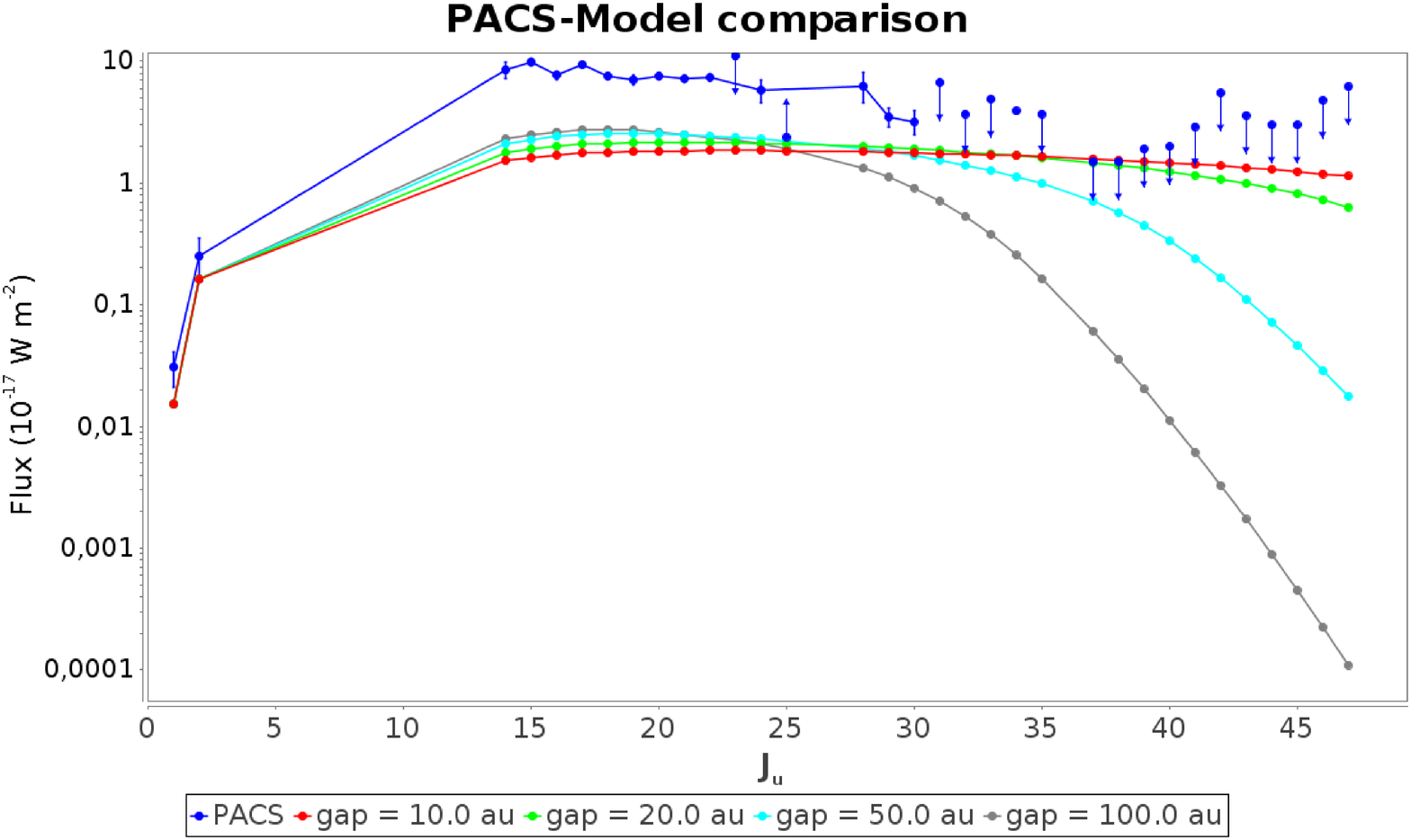}
   \caption{Comparison between PACS observations and our radiative transfer model for different values of the 
   inner CO gap, and assuming a standard $^{12}$CO abundance (1.6$\times$10$^{-4}$). The values 
   for J$_{u}$ of 1 and 2 are those found by \cite{Fuente2006}.}
    \label{fig:PACS_model}%
\end{figure*}

In Fig. \ref{fig:PACS_model} we compare the  PACS fluxes with our model calculations.
Thirty-one CO lines with J$_u$$>$14 have been observed using PACS. As expected,
our model predicts that all these lines are optically thick in
the inner disk and only become optically thin at large radii. 
For a face-on disk, the opacities of the CO J=14$\rightarrow$13, 19$\rightarrow$18,
and 39$\rightarrow$38 lines are lower than 1 for radii larger than $\sim$900 au, $\sim$400 au, and
$\sim$100 au, respectively. Since $\sim$80\% of the emission comes from the optically thick part, different 
CO abundances and/or small changes of the disk mass
only translates into slight differences in the total line fluxes. In addition, the
assumed R$_{out}$ has a negligible impact on the J$_u$$>$30 PACS fluxes as long as 
R$_{out}$$>$100 au. \citet{Alo09} obtained an outer radius of 150$\pm$50~AU for the dusty disk. 
Since gaseous disks are usually larger than dusty disks \citep{Facchini17}, we 
can reasonably assume that the radius of the gaseous disk is larger than this value. 
  
Two different regions are clearly differentiated in the CO SED shown
in Fig. \ref{fig:PACS_model}.
Our model falls short of accounting for the fluxes of the mid-J (J$_u$=14-30) CO lines. 
Since the lines are optically thick,  increasing the disk gas mass or the CO abundance 
does not change our results but an increase in the turbulent velocity would certainly translate 
into an increase in the total line flux. The only way to push up the estimated fluxes to the observed 
values is to consider a turbulent velocity of $\sim$ 5 km s$^{-1}$. 
However, this value is not consistent with the interferometric
CO 1$\rightarrow$0 and 2$\rightarrow$1 profiles. It is also higher than the 
turbulent velocities observed in other Herbig Ae and T Tauri disks (see, e.g., \citealp{Pietu2007}). 
Therefore, we consider that the high J$_u$=14$-$30 CO line fluxes are better explained by 
some contribution 
from the envelope and/or outflow to the emission of these lines \citep{JimenezD2017}. 
Taking into account that these lines are unresolved with the PACS spectral and spatial
resolution, we favor the interpretation of some contribution from the inner envelope
and/or the UV-illuminated walls of the molecular outflow to the J$_u$=14-30 lines \citep{Karska2014}. We recall that \citet{Alo09} 
needed to assume a warm envelope
to fit the FIR part of the continuum SED.

In contrast, the fluxes of the J$_u$$>$31 CO lines are slightly overestimated by our model
for R$_{in}$$\leq$20 au. This suggests the existence of a molecular cavity in the R Mon disk.
This result is not unexpected since \citet{Alo09} determined 
an inner radius of $\sim$18 au for the dusty disk. Furthermore, the continuum SED
toward R Mon is better fitted without an inner rim. An optically thick CO disk is
not expected in the region depleted in dust \citep{Banzatti2017}.
We propose that the disk around the early Be star R Mon is a transition disk 
by a large inner region depleted in dust and molecular gas, i.e., a low-NIR
cavity following the nomenclature of \cite{Banzatti2017}. 
We are aware, however, of the limitations and uncertainties of our model, which are discussed
in detail in the next section. 


%
%

\section{Uncertainties in our model}
\label{uncertainties}

\begin{figure}
    \centering
    \includegraphics[width=0.5\textwidth]{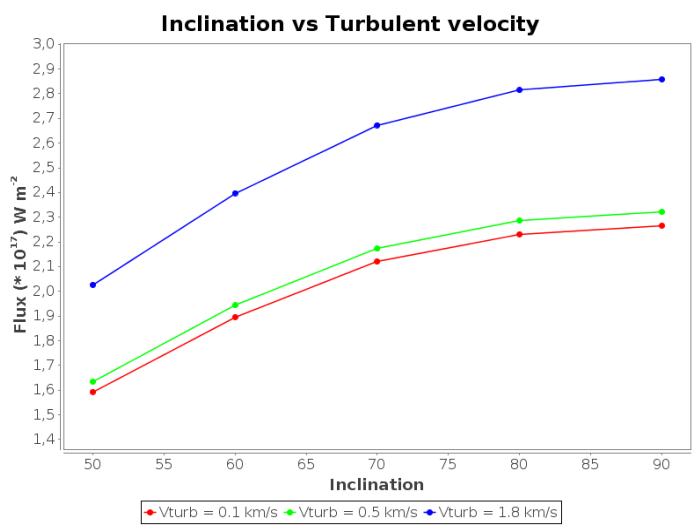}
   \caption{Results of our CO model for line J=37$\rightarrow$36, calculated for different values of the inclination of the disk and turbulent velocity.} 
    \label{fig:X}%
\end{figure}

Our estimate of the inner radius of the gaseous disk is based on
the upper limits to the high-J CO lines that are expected to be optically thick for R$<$100 au. 
The total fluxes are then 
proportional to the gas temperature and the observed line width, that is itself dependent on the disk 
inclination (because of the Keplerian rotation), gas temperature (thermal dispersion), and turbulent velocity. 
The key parameter in this result is the value of T$_{in}$, which is model dependent. 
In our simple approximation, we assumed that the gas temperature is the 
same as the dust temperature in the blackbody limit. For each inner radius,
this value depends on the assumed stellar radius and effective temperature. For this reason, it is essential to have an accurate
estimate of the the stellar spectral type for our discussion. 

But even for a given set of stellar parameters, it is not easy
to determine the precise value of the molecular gas temperature in the inner edge of the disk.
In principle, it depends on the disk morphology (flaring), grain properties, dust settling, and on the distance from the star. In addition, the cavity could be filled with a tenuous gas and dust
that could partially shield the stellar radiation, leading to a lower T$_{in}$ for the same inner radius.
We decided to adopt the blackbody limit to estimate T$_{in}$, which is a conservative value.
In the illuminated disk surface, the dust is expected to be super-heated to
temperatures higher than the blackbody limit \citep{Dullemond01}. Moreover, the photoelectric effect is an 
efficient gas heating mechanism in the illuminated surfaces where the  gas temperature is 
expected to be higher than the dust temperature.
A more realistic model with detailed thermal balance is expected to produce higher line fluxes
and consequently, we get a higher T$_{in}$ for a given R$_{in}$ and hence, we would need to assume a larger cavity to fit PACS observations.

Although less important, the adopted disk inclination and turbulent velocity also have an impact on 
the predicted size of the disk cavity. 
In Fig. \ref{fig:X}  we show the total flux of the J=37$\rightarrow$36 line as a function of inclination and turbulent velocity.
The inclination uncertainty has a minor effect because the smaller projected area of the disk at high 
inclinations is to some extent compensated by the higher velocity gradient along the line of sight. Turbulent velocity
is more critical, especially in those regions in which it is comparable to the thermal dispersion. 
We recall that the value used in our model, $\sim$0.5 km s$^{-1}$, has been 
derived from the fitting of the interferometric $^{12}$CO 1$\rightarrow$0 and 2$\rightarrow$1 line profiles. 
Assuming a lower value of $\sim$0.1 km s$^{-1}$, similar to that found in low-mass and Herbig Ae stars by \cite{Pietu2007}, 
we would derive an inner radius $>$10 au to account for the upper limit to the J=37$\rightarrow$36 line emission.

\section{Molecular chemistry: $^{13}$CO, CN, HCO$^+$, H$_2$O}
\label{section:Chem}

\begin{figure*}
    \centering
    \includegraphics[width=1\textwidth]{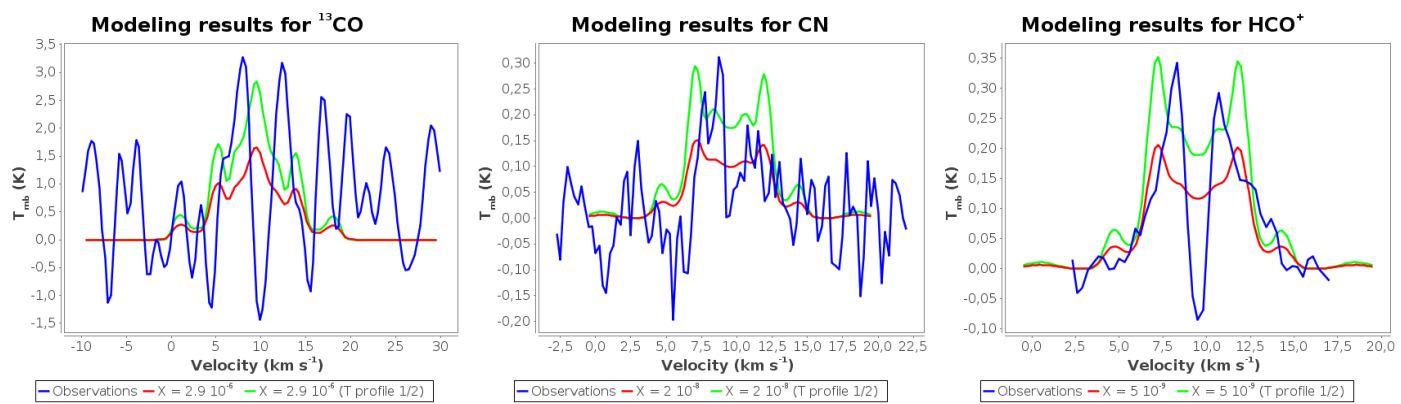}
   \caption{Modeling results for the temperature profile assumed in the $^{12}$CO model and for a second profile 
   equal to half of that. $^{13}$CO, CN, and HCO$^+$ are shown from left to right.} 
    \label{fig:HCOp_CN_13CO_model}%
\end{figure*}

We have used a reference model with R$_{in}$=50 au to fit the PdBI observations of $^{13}$CO 1$\rightarrow$0 line.
We estimate for R Mon a $^{12}$C/$^{13}$C isotopic ratio of 55$\pm$15 \citep{Savage2002}. 
Because of their different optical depths,
the emission of the $^{13}$CO 1$\rightarrow$0 line is expected to come from a layer closer to 
the midplane, and thus colder, than the layer emitting in $^{12}$CO 1$\rightarrow$0. 
Based on the results of \citet{Pietu2007}, we assume that the $^{13}$CO 1$\rightarrow$0 line is thermalized and 
the excitation temperature is half of that corresponding to $^{12}$CO, reaching a minimum 
value around 25 K in the external region of the disk. The resulting spectrum is plotted 
in Fig. \ref{fig:HCOp_CN_13CO_model}, left panel. Because of the low S/N ratio of our
$^{13}$CO 1$\rightarrow$0 observations and the spatial filtering of the emission in the velocities close to that of the molecular cloud, the two predicted profiles are consistent with our observations.

For CN and HCO$^+$ we used the same model that successfully fit the $^{12}$CO and $^{13}$CO data to fit the CN and HCO$^+$
observations. In these calculations we assume the same excitation temperature as in the case of the
$^{13}$CO 1$\rightarrow$0 line and vary the CN and HCO$^+$ abundance to fit the observed spectrum. 
As a first approximation, we assume that the molecular abundances are uniform along the disk.
In Fig. \ref{fig:HCOp_CN_13CO_model}, we show the predicted profiles assuming a fractional abundance (wrt H$_2$) of 2 $\times$ 10$^{-8}$ for CN 
and 5 $\times$ 10$^{-9}$ for HCO$^{+}$. The line profiles predicted by our model are fairly consistent with the
observations taking the limitations of our approximation into account.

These abundances are a factor of $\sim$10 larger than the averaged molecular abundances derived 
in T Tauri and Herbig Ae disks. However, these abundances are similar to those
found in photon-dominated regions (see \citealp{Ginard2012} and references therein).
CN has been widely used as tracer of photon-dominated regions in different environments including reflection nebulae \citep{Fuente1993,Fuente1995},
proto-planetary nebulae \citep{Bachiller1997}, HII regions \citep{Fuente1996,Rodriguez1998,Ginard2012}, and external 
galaxies \citep{Fuente2005}. While the CN/HCO$^+$$\sim$1 in dark clouds, a value of $\sim$3 is found in the 
Orion Bar, very similar to that in the R Mon disk. 
We are aware that both the absolute value of the
abundances and the CN abundance relative to HCO$^+$ are very uncertain because we are using a simple disk model that 
neglects the density and temperature variations in the vertical direction.
Nevertheless, the good agreement between the estimated values with those found in 
the surface of photon-dominated regions provides further support to the important role played by the stellar UV radiation in 
the disk evolution.

It is also interesting to comment on the lack of a clear detection of any H$_2$O line in the PACS spectrum. Within the GASPS project,
the water line at $\sim$ 63~$\mu$m has been detected in around 10\% of the observed T Tauri \citep{Riviere2012} and Herbig 
Ae/Be \citep{Fedele2012} stars. These lines are interpreted as coming from warm water (T$_k$$>$200 K) in the inner disk
region. The nondetection of these high-excitation H$_2$O lines are also consistent with the presence of a cavity in the
circumstellar disk around R Mon.



\section{Discussion}
\label{section:Comparison}
Planet formation occurs in protoplanetary disks during the early stages of stellar evolution. These
protoplanetary systems evolve from massive disks with a typical gas-to-dust ratio of 100 to optically
thin systems similar to our Kuiper Belt. The evolution of the disk is governed by viscous transport
\citep{Muzerolle1998} and photoevaporation \citep{Hollenbach1994,Hollenbach2000,Clarke2001} 
and models combining both had been successfully used to explain
the dispersal of protoplanetary disks around low-mass TTS \citep{Alexander2006a,Alexander2006b}. In the so-called UV-switch
model, evolution proceeds mostly through viscous accretion to a point where most accretion stops.
From this point on, the gas is quickly dispersed by photoevaporation. This is known as the two
timescale problem. Also, planet formation is considered to play a major role in cleaning the inner
parts of the disk in the early evolution of these systems \citep{Zhu2012}.
To what extent the same mechanisms can be used to explain the evolution of more massive Herbig Ae/Be
stars remains uncertain. Even if mechanisms driving the evolution are the same, we still need
to understand the contribution of each of these mechanisms due to the very different mass regimes involved.
For instance, photoevaporation is expected to be much more important in Herbig Ae/Be stars than
in the low-mass regime. 

\citet{Alo09} proposed photoevaporation as the main dispersal mechanism in disks around massive stars. Since 
the ionizing flux in Herbig Be stars is one order of magnitude higher than in the cooler Herbig Ae and TTS, 
the timescale for disk photoevaporation is shorter and inner gaps are expected to be formed in a few 0.1~Myr.
The characteristic gap radius is the so-called gravitational radius, defined as the radius at which the thermal velocity is
equal to the escape velocity and the hot gas overcomes the gravitational field of the star. For a B0 star such as R~Mon, the gravitational
radius is $\sim$70 au \citep{Alexander2006b}. We used Herschel/PACS spectroscopic observations combined with 
interferometric observations of the low-J lines of CO and $^{13}$CO to derive
the size of the inner cavity in the R Mon disk.  The upper limits of the J$_u$$>$31 CO lines 
as measured with PACS are consistent with the existence of an inner cavity with a radius similar to or larger than of
that of the dusty disk. The PDR-like chemistry associated with this disk also supports the great influence of the UV radiation
in the disk evolution. 



\section{Summary and conclusions}
\label{section:Summary}
   Our goal is to combine Herschel/PACS data with our interferometric observations at millimeter wavelengths
to have a deeper insight into the physics and chemistry of the R Mon disk.
Interferometric detections of the HCO$^+$ 1$\rightarrow$0, and CN 1$\rightarrow$0 lines using the IRAM 
Plateau de Bure Interferometer (PdBI) are presented as well. The HCN 1$\rightarrow$0 line was searched but not detected.
In order to constrain the disk model, we re-evaluated the R Mon spectral type using the stellar SED 
and UVES emission lines. Our analysis confirms that R Mon is a B0 star. Our disk model falls short of predicting the fluxes of 
the 14$<$J$_u$$<$31 lines. Emission coming from the remnant envelope  and/or from the UV-illuminated walls of the bipolar outflow
could contribute toward increasing the observed value of these lines to some extent. The upper limits of the J$_u$$>$31 CO lines 
as measured with PACS are consistent with the existence of an inner cavity with a radius similar or larger than that of
the dusty disk. The intense emission of the HCO$^+$ and CN lines suggest a strong influence of the UV photons on the 
gas disk chemistry as well. All the observations gathered thus far are hence consistent with the scenario of 
R Mon being a transition disk with a cavity of R$_{in}$$\gtrsim$20 au. This size is consistent with the photoevaporation radius
suggesting that UV photoevaporation could be the main disk dispersal mechanism in massive disks.
However, the limited angular resolution of our observations, the 
absence of velocity resolved profiles of the high-J lines, and the large number of unknown variables
in the disk modeling preclude an undoubtful conclusion.

\begin{acknowledgements}
We acknowledge the Spanish MINECO for
funding support under grants AYA2012-32032.
AYA2016-75066-C2-1/2-P and ERC under ERC-2013-SyG, G. A. 610256 NANOCOSMOS.
BM is supported by Spanish grant AYA 2014-55840-P.
Based on data products from observations made with ESO Telescopes at
the Paranal Observatory under program ID 082.C-0831 (PI: Mario
Van den Ancker). 
SPTM acknowledges to the European Union's Horizon 2020 
research and innovation program for funding support given 
under grant agreement No~639459 (PROMISE).
\end{acknowledgements}

\bibliography{RMon_v2}
\bibliographystyle{aa}

\newpage
\appendix

\section{Figures and Tables}

\begin{figure}
    \centering
    \includegraphics[width=0.5\textwidth]{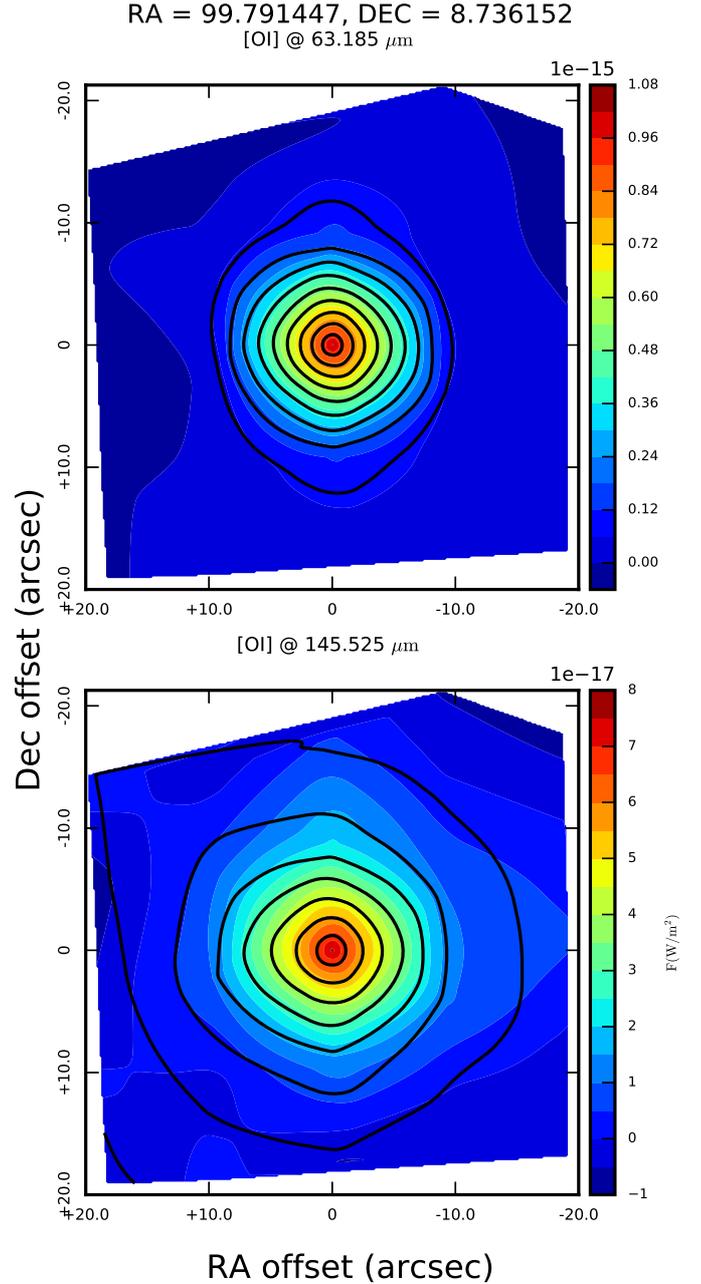}
    \caption{PACS maps for [OI] lines at 63 and 145 $\rm \mu m$. }
    \label{fig:PACS_OImaps}
\end{figure}

\begin{figure*}
    \centering
    \includegraphics[width=1.0\textwidth]{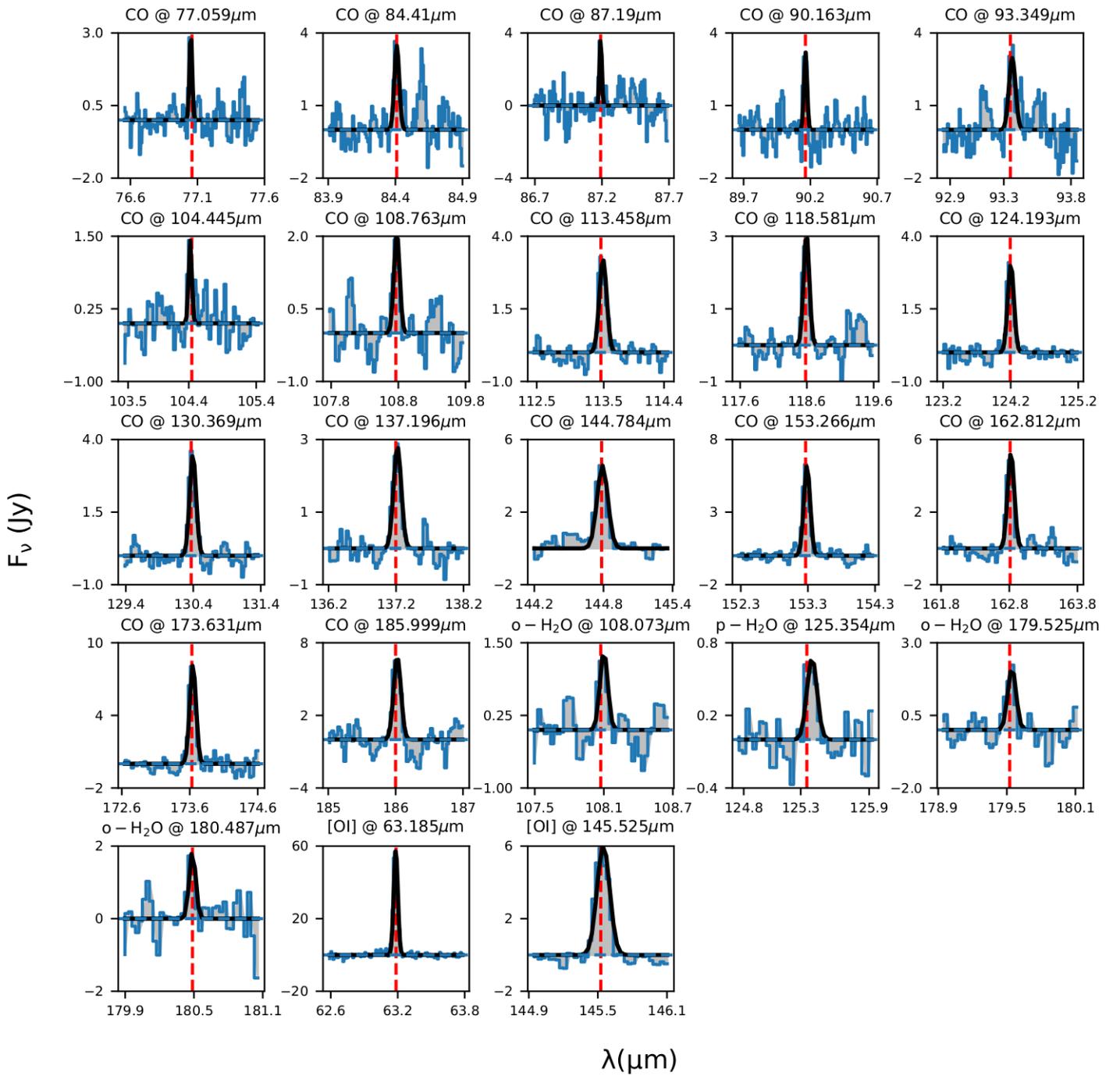}
    \caption{PACS emission lines and Gaussian fittings for the transitions detected in the 
    central spaxel. }
    \label{fig:PACS_detectedLines}
\end{figure*}

\begin{table}
\caption{PACS $\rm H_{2}O$ line fluxes from the central spaxel.}           
\label{table:PACSCO_H2O_fluxes} 
\centering                      
\begin{tabular}{crc}       
\hline\hline
Transition       & \multicolumn{1}{c}{$\lambda$}         & F \\
                      & \multicolumn{1}{c}{($\rm \mu m$)}    & $\rm (10^{-17} W/m^{2}$)\\ 
\hline
$\rm H_{2}O$ & 56.325 &  $\rm <$ 1.9 \\
$\rm H_{2}O$ & 56.816 &  $\rm <$ 2.0 \\
$\rm H_{2}O$ & 57.637 &  $\rm <$ 1.6 \\
$\rm H_{2}O$ & 58.699 &  $\rm <$ 1.2 \\
$\rm H_{2}O$ & 59.987 &  $\rm <$ 1.2 \\
$\rm H_{2}O$ & 63.323 &  $\rm <$ 6.7 \\
$\rm H_{2}O$ & 63.457 &  $\rm <$ 7.5 \\
$\rm H_{2}O$ & 65.166 &  $\rm <$ 1.3 \\
$\rm H_{2}O$ & 66.093 &  $\rm <$ 1.2 \\
$\rm H_{2}O$ & 67.089 &  $\rm <$ 0.9 \\
$\rm H_{2}O$ & 67.269 &  $\rm <$ 0.9 \\
$\rm H_{2}O$ & 71.067 &  $\rm <$ 3.1 \\
$\rm H_{2}O$ & 71.540 &  $\rm <$ 2.9 \\
$\rm H_{2}O$ & 71.947 &  $\rm <$ 2.5 \\
$\rm H_{2}O$ & 78.742 &  $\rm <$ 1.8 \\
$\rm H_{2}O$ & 78.928 &  $\rm <$ 1.8 \\
$\rm H_{2}O$ & 84.767 &  $\rm <$ 1.7 \\
$\rm H_{2}O$ & 89.988 &  $\rm <$ 1.2 \\
$\rm H_{2}O$ & 108.073 & 3.4 $\rm \pm$ 1.0 \\
$\rm H_{2}O$ & 111.628 &  $\rm <$ 1.2 \\
$\rm H_{2}O$ & 113.948 &  $\rm <$ 1.2 \\
$\rm H_{2}O$ & 121.722 &  $\rm <$ 0.7 \\
$\rm H_{2}O$ & 125.354 & 2.1 $\rm \pm$ 0.6 \\
$\rm H_{2}O$ & 126.714 & $\rm <$ 1.5  \\
$\rm H_{2}O$ & 127.884 & $\rm <$ 0.8 \\
$\rm H_{2}O$ & 132.408 & $\rm <$ 0.7  \\
$\rm H_{2}O$ & 134.935 & $\rm <$ 1.1  \\
$\rm H_{2}O$ & 136.496 & $\rm <$ 0.6  \\
$\rm H_{2}O$ & 138.528 & $\rm <$ 0.9  \\
$\rm H_{2}O$ & 144.517 & $\rm <$ 0.5  \\
$\rm H_{2}O$ & 158.312 & $\rm <$ 0.7  \\
$\rm H_{2}O$ & 179.525 &  1.4 $\rm \pm$ 0.4  \\
$\rm H_{2}O$ & 180.487 &  1.1 $\rm \pm$ 0.3  \\
\hline                                   
\end{tabular}
\end{table}

\begin{figure*}
    \centering
    \includegraphics[width=1.0\textwidth]{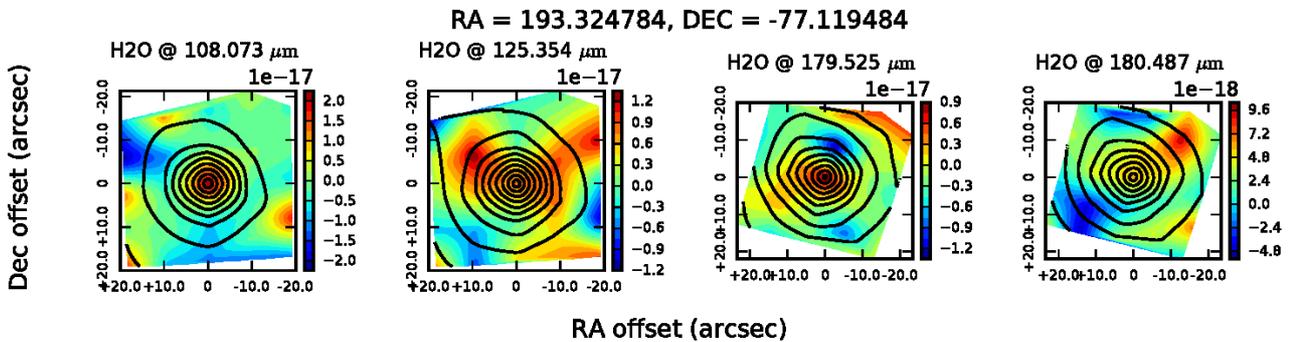}
    \caption{PACS maps for  detected $\rm H_{2}O$ lines.  }
    \label{fig:PACS_H2Omaps}
\end{figure*}

\begin{figure*}
    \centering
    \includegraphics[width=1.0\textwidth]{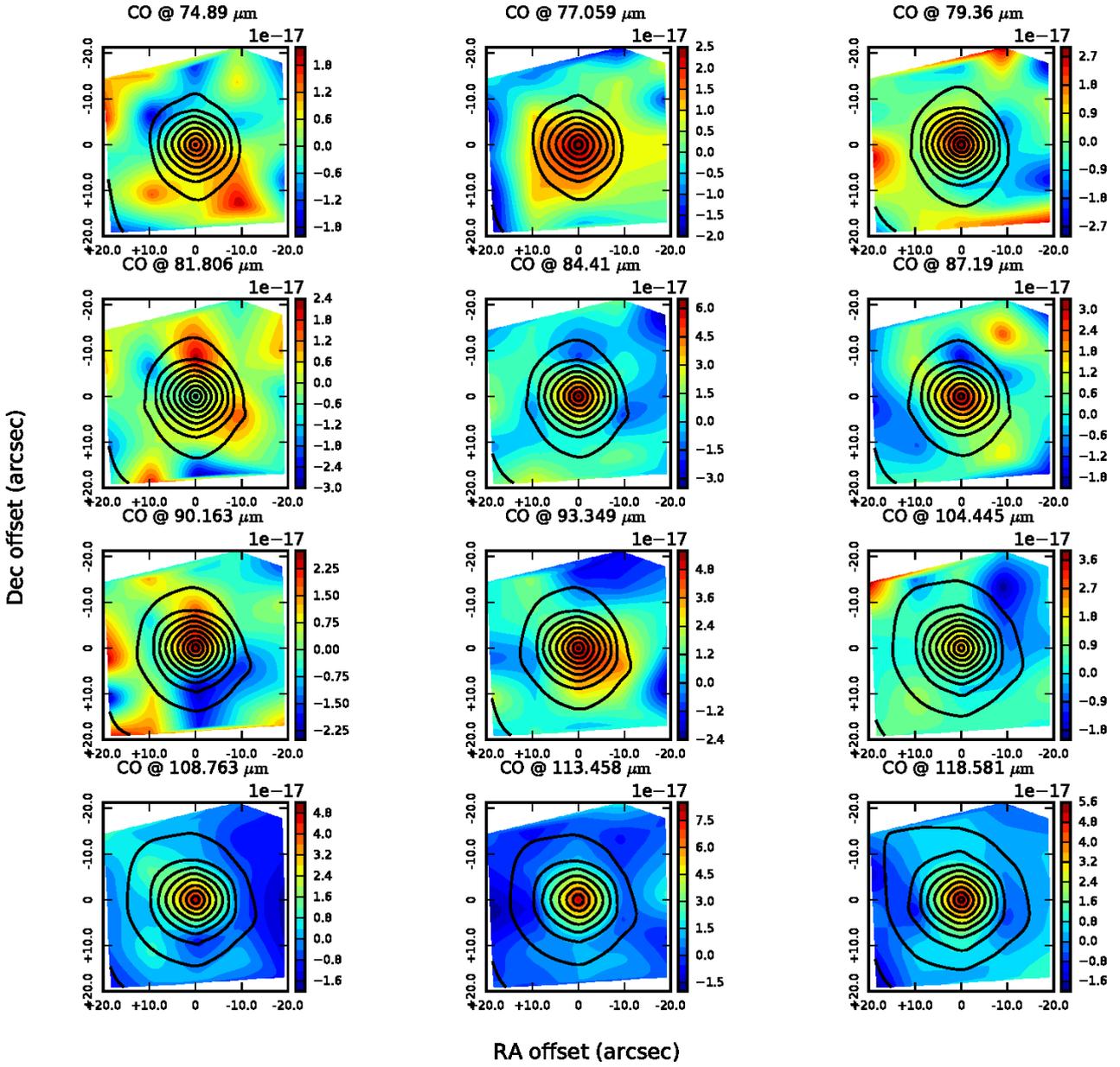}
    \caption{PACS maps for CO lines in the 74-119 $\rm \mu m$ range ($\rm J_{up}$ 35 to 22). }
    \label{fig:PACS_COmaps_1}
\end{figure*}

\begin{figure*}
    \centering
    \includegraphics[width=1.0\textwidth]{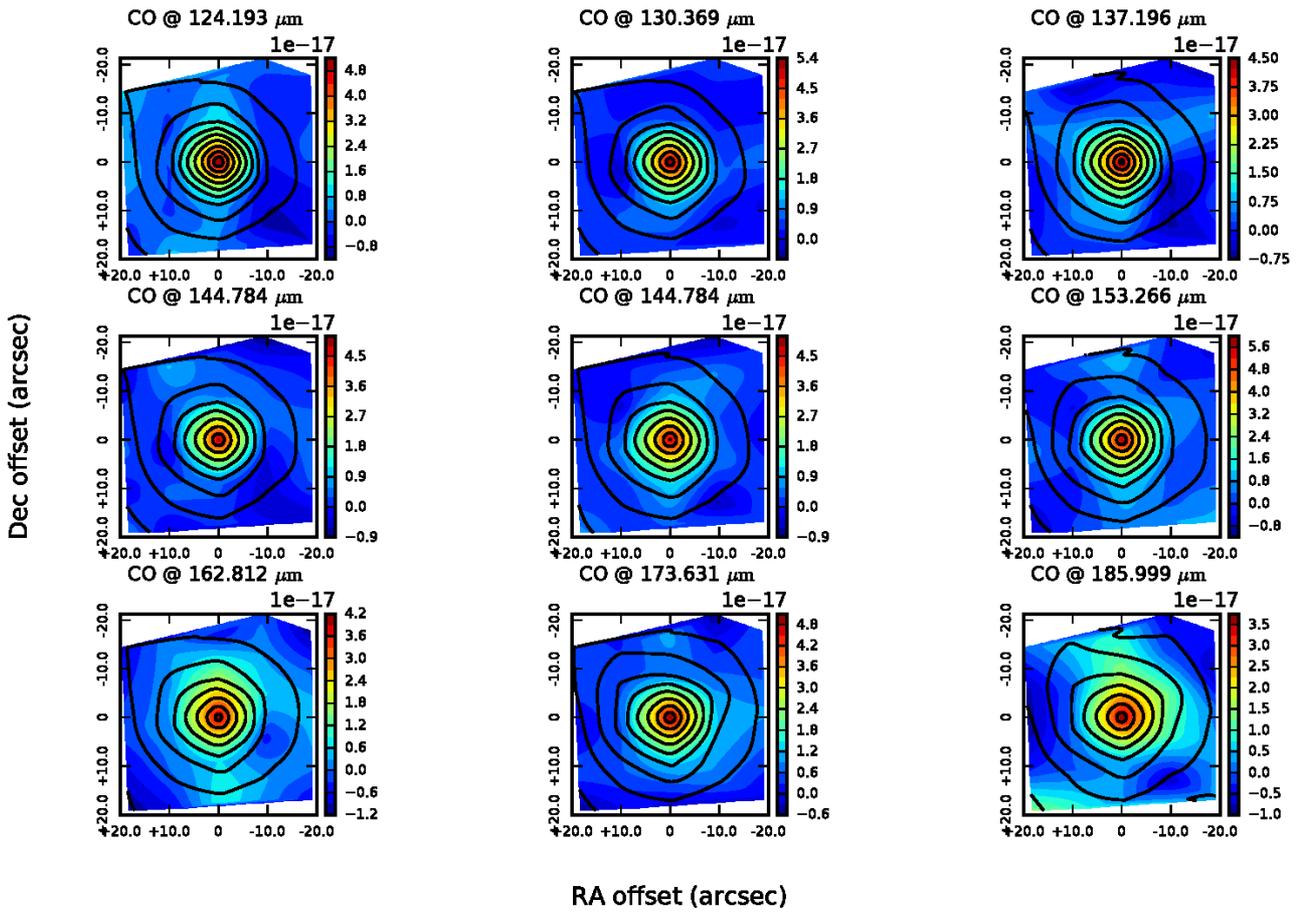}
    \caption{PACS maps for CO lines in the 124-186 $\rm \mu m$ range ($\rm J_{up}$ 21 to 14).  }
    \label{fig:PACS_COmaps_2}
\end{figure*}

\end{document}